\documentclass[12pt]{article}
\usepackage{epsfig}\parskip 5pt plus 1pt
\usepackage{bbm}
\usepackage{amsmath}
\usepackage{amssymb}
\usepackage{amsfonts}
\usepackage{mathrsfs}
\usepackage{graphicx}
\usepackage{axodraw}
\usepackage{color}
\newcommand{\nn}{\nonumber}

\newcommand{\be}{\begin{equation}}
\newcommand{\ee}{\end{equation}}
\newcommand{\bea}{\begin{eqnarray}}
\newcommand{\eea}{\end{eqnarray}}

\begin{document}
\begin{titlepage}
\vspace*{-2cm}
\flushright{MPP-2008-59}
\vskip 1.5cm
\begin{center}
{\Large\bf The contribution of fermionic seesaws}\\[.3cm]
{\Large\bf to the anomalous magnetic moment of leptons}
\end{center}
\vskip 0.5cm
\begin{center}
{\large C.~Biggio}~\footnote{biggio@mppmu.mpg.de}
\vskip 0.5cm
Max-Planck-Institut f\"ur Physik, 80805 M\"unchen, Germany
\end{center}
\vskip 0.5cm

\begin{abstract}
We calculate the contribution to the anomalous magnetic moment of
leptons in the type-I and type-III seesaw models. We show that, even
if the scale of the new physics is pushed down to the electroweak
scale, this contribution is not large enough to explain the measured
discrepancy of the muon anomalous magnetic moment.
\end{abstract}
\end{titlepage}
\setcounter{footnote}{0}
\vskip2truecm

\section{Introduction}
\label{intro}

The anomalous magnetic moment of a spin-1/2 particle is defined as the
difference between the gyromagnetic factor $g$ of that particle and
its value in the Dirac theory. In formulae, $a=(g-2)/2$, with $g$
defined by $\vec{\mu}=g\frac{e}{2m}\vec{s}$, where $\vec{\mu}$ is the
particle's magnetic moment, $\vec{s}$ its spin and $e$ and $m$ its
electric charge and mass. Deviations from the Dirac's value $g=2$ are
predicted in quantum theories due to loops effects.

The measurements of the electron and muon anomalous magnetic moments
have recently reached an extraordinary precision, which has permitted,
through the former, to give the most precise determination of the
fine-structure constant $\alpha$ and, using the latter, to reveal a
discrepancy between the standard model (SM) value and the experimental
one. This discrepancy, that could be due to new physics beyond the SM,
has motivated the calculation of $a_\mu$ in many different models like
models with supersymmetry, extra dimensions, extended Higgs sector,
additional gauge
bosons...~\cite{Czarnecki:2001pv,Stockinger:2006zn}. Since new physics
is required to explain neutrino masses, one may wonder if this could
also affect observables like the anomalous magnetic moment. Even if it
is usually believed that the new physics responsible for neutrino
masses lies at a very high energy scale and therefore contributes in a
negligible way to these low energy processes, it is not excluded that
it can live at the energies which will be reached at near-future
colliders like LHC. If this is the case, the contribution to low
energy processes could be relevant and it is therefore interesting to
study these effects.

The theoretical prediction of the electron anomalous magnetic moment
is dominated by QED contributions which have been calculated up to
four loops~\footnote{See Ref.~\cite{Passera:2007fk} for a short review
  on $e$, $\mu$ and $\tau$ anomalous magnetic moment and references
  therein.}~\cite{Schwinger:1948iu}. Adding to them the electroweak
(EW) and the hadronic contributions~\cite{altri} and comparing the
result with the measured value~\cite{Odom:2006zz}
\be
\label{aeexp}
a_e^{exp}=115\ 965\ 218\ 0.85 \ (76) \cdot 10^{-12}\, ,
\ee
the most precise determination of $\alpha$ has been derived:
\be
\label{alpha}
\alpha^{-1}\ =\ 137.035\ 999\ 709\ (96)\, .
\ee

The present world average experimental value of the muon anomalous
magnetic moment is given by~\cite{Bennett:2004pv}
\be
\label{amuexp}
a_\mu^{exp}=116\ 592\ 080\ (63) \cdot 10^{-11}\, .
\ee
On the theoretical side the reached precision is almost the same,
where the error is dominated by the hadronic contribution. This can be
extracted either from the hadronic $e^+e^-$ annihilation data or from
hadronic $\tau$ annihilation data. Depending on which of them is
considered, the $a_\mu^{SM}$ ranges between $116\ 591\ 748\ (61)\cdot
10^{-11}$~\cite{Hagiwara:2006} and $116\ 591\ 961\ (70)\cdot
10^{-11}$~\cite{DEHZ03}, which give rise to the following
discrepancies~\footnote{Intermediate values have been found in
  Refs.~\cite{intermedimu}.}:
\bea
\label{deltaamuhigh}
\delta a_\mu &=& 332\ (88)\cdot 10^{-11} \qquad (3.8\ \sigma)\\
\label{deltaamulow}
\delta a_\mu &=& 119\ (95)\cdot 10^{-11} \qquad (1.3\ \sigma)\, .
\eea
Roughly, the difference between the SM prediction and the measured
value is of $\mathcal{O}(10^{-9})$. Any model of new physics able to
contribute to the muon anomalous magnetic moment by a similar amount
could be a good candidate to explain this discrepancy.

As for the tau anomalous magnetic moment, the experimental
precision~\cite{Abdallah:2003xd} is unfortunately too low compared
with the theoretical one~\cite{Passera:2007fk}, so that no useful
information can be obtained so far:
\be
\label{atauexp}
-0.052 < a_\tau^{exp} < 0.013
\ee
\be
\label{atauSM}
a_\tau^{SM}=117\ 721\ (5) \cdot 10^{-8}\, .
\ee

In this paper we calculate the contribution to the anomalous magnetic
moment of the leptons due to singlets (triplets) of fermions
responsible for generating neutrino masses through the type-I (III)
seesaw mechanism~\cite{TIPOI} (\cite{Foot:1988aq}). The organization
of this paper is the following: in Sect.~\ref{typeIII} we define our
notations and calculate the anomalous magnetic moment due to triplets
of fermions, in Sect.~\ref{typeI} we derive the result for the
singlets case, while Sect.~\ref{conclu} contains our conclusions.

\section{The anomalous magnetic moment in the type-III seesaw model}
\label{typeIII}

\subsection{The type-III seesaw Lagrangian}
\label{lagrangiantypeIII}

The type-III seesaw model consists in the addition to the SM of SU(2)
triplets of fermions with zero hypercharge, $\Sigma$.  In this model
at least two such triplets are necessary in order to have two
non-vanishing neutrino masses, while there can be a contribution to
the anomalous magnetic moment even with only one triplet. Therefore,
in the following, we will not specify the number of triplets and our
result will be independent of it. Following the notations of
Ref.~\cite{Abada:2008ea}, the beyond-the-SM Lagrangian is given by
\begin{equation}
\label{Lfermtriptwobytwo}
{\cal L}=Tr [ \overline{\Sigma} i \slash \hspace{-2.5mm} D  \Sigma ] 
-\frac{1}{2} Tr [\overline{\Sigma}  M_\Sigma \Sigma^c 
                +\overline{\Sigma^c} M_\Sigma^* \Sigma] 
- \tilde{\phi}^\dagger \overline{\Sigma} \sqrt{2}Y_\Sigma L 
-  \overline{L}\sqrt{2} {Y_\Sigma}^\dagger  \Sigma \tilde{\phi}\, ,
\end{equation} 
where $L\equiv (l,\nu)^T$, $\phi\equiv (\phi^+, \phi^0)^T\equiv
(\phi^+, (v+H+i \eta)/\sqrt{2})^T$ with $v\equiv \sqrt{2} \langle
\phi^0 \rangle=246$~GeV, $\tilde \phi = i \tau_{2} \phi^*$, $\Sigma^c
\equiv C \overline{\Sigma}^T$ and where, for each fermionic triplet,
\be
\Sigma=
\left(
\begin{array}{ cc}
   \Sigma^0/\sqrt{2}  &   \Sigma^+ \\
     \Sigma^- &  -\Sigma^0/\sqrt{2} 
\end{array}
\right) \, .
\ee
Without loss of generality, in the following we will assume that we
start from the basis where $M_\Sigma$ is real and diagonal. We express
the four degrees of freedom of each charged triplet in terms of a
single Dirac spinor $\Psi\equiv\Sigma_R^{+ c} + \Sigma_R^-$, so that
the field content of our model is given by the SM fields $plus$ $n$
Majorana fermions $\Sigma^0$ (right-handed neutrinos) $plus$ $n$
charged Dirac fermions $\Psi$. In the mass basis, the Lagrangian terms
relevant to our calculation are the following
\begin{equation}
\mathcal{L}\ \supset\
\mathcal{L}_{CC}+\mathcal{L}_{NC}+
\mathcal{L}_{H,\eta}+\mathcal{L}_{\phi^{-}}\, ,
\label{fulllagrangian}
\end{equation}
where $\mathcal{L}_{CC}$ contains the charged current interactions
among the charged leptons $l$, the charged fields $\Psi$, the
neutrinos $\nu$ and the neutral fields $\Sigma^0$, $\mathcal{L}_{NC}$
are the neutral current interactions for charged fermions only and
$\mathcal{L}_{H,\eta}$ and $\mathcal{L}_{\phi^{-}}$ contain the
interactions of the fermions with the Higgs and the Goldstone bosons
(see Appendix~A for details). In this model the neutrino mass matrix
is given by:
\begin{equation}
\label{mnu}
m_\nu=-\frac{v^2}{2} Y_\Sigma^T\frac{1}{M_\Sigma} Y_\Sigma\,.
\end{equation}
%

\begin{figure}[t]
\centering
\vspace{15mm}
\begin{picture}(330,150)(0,-100)
\ArrowLine(10,80)(50,80)
\ArrowLine(50,80)(90,80)
\ArrowLine(90,80)(130,80)
\DashCArc(70,80)(20,0,180){3}
\Photon(70,100)(70,130){1.5}{5}
\Text(10,82)[bl]{$l$}
\Text(130, 82)[br]{$l$}
\Text(74,130)[tl]{$\gamma$}
\Text(57,95)[br]{\small{$\phi^-$}}
\Text(88,95)[bl]{\small{$\phi^-$}}
\Text(70,82)[b]{\scriptsize{$\nu,\Sigma^0$}}
\ArrowLine(180,80)(220,80)
\ArrowLine(220,80)(260,80)
\ArrowLine(260,80)(300,80)
\DashCArc(240,80)(20,90,180){3}
\PhotonArc(240,80)(20,0,90){2}{5.5}
\Photon(240,100)(240,130){1.5}{5}
\Text(180,82)[bl]{$l$}
\Text(300, 82)[br]{$l$}
\Text(244,130)[tl]{$\gamma$}
\Text(227,95)[br]{\small{$\phi^-$}}
\Text(258,95)[bl]{\small{$W^-$}}
\Text(240,82)[b]{\scriptsize{$\nu,\Sigma^0$}}
\ArrowLine(10,0)(50,0)
\ArrowLine(50,0)(90,0)
\ArrowLine(90,0)(130,0)
\DashCArc(70,0)(20,0,90){3}
\PhotonArc(70,0)(20,90,180){2}{5.5}
\Photon(70,20)(70,50){1.5}{5}
\Text(10,2)[bl]{$l$}
\Text(130, 2)[br]{$l$}
\Text(74,50)[tl]{$\gamma$}
\Text(57,15)[br]{\small{$W^-$}}
\Text(88,15)[bl]{\small{$\phi^-$}}
\Text(70,2)[b]{\scriptsize{$\nu,\Sigma^0$}}
\ArrowLine(180,0)(220,0)
\ArrowLine(220,0)(260,0)
\ArrowLine(260,0)(300,0)
\PhotonArc(240,0)(20,0,180){2}{10.5}
\Photon(240,22)(240,50){1.5}{5}
\Text(180,2)[bl]{$l$}
\Text(300,2)[br]{$l$}
\Text(244,50)[tl]{$\gamma$}
\Text(227,15)[br]{\small{$W^-$}}
\Text(258,15)[bl]{\small{$W^-$}}
\Text(240,2)[b]{\scriptsize{$\nu,\Sigma^0$}}
\ArrowLine(10,-60)(50,-60)
\ArrowLine(50,-60)(70,-60)
\ArrowLine(70,-60)(90,-60)
\ArrowLine(90,-60)(130,-60)
\PhotonArc(70,-60)(20,0,180){2}{10.5}
\Photon(70,-60)(70,-85){1.5}{5}
\Text(10,-59)[bl]{$l$}
\Text(130, -59)[br]{$l$}
\Text(74,-85)[bl]{$\gamma$}
\Text(70,-37)[b]{\small{$Z$}}
\Text(70,-59)[b]{\scriptsize{$l',\Psi$}}
\ArrowLine(180,-60)(220,-60)
\ArrowLine(220,-60)(240,-60)
\ArrowLine(240,-60)(260,-60)
\ArrowLine(260,-60)(300,-60)
\DashCArc(240,-60)(20,0,180){3}
\Photon(240,-60)(240,-85){1.5}{5}
\Text(180,-59)[bl]{$l$}
\Text(300, -59)[br]{$l$}
\Text(244,-85)[bl]{$\gamma$}
\Text(240,-37)[b]{\small{$H,\eta$}}
\Text(240,-59)[b]{\scriptsize{$l',\Psi$}}
\end{picture}
\caption{Diagrams contributing to the anomalous magnetic moment of the
  lepton $l$. $\phi^\pm,\, \eta$ are the three Goldstone bosons
  associated with the $W^\pm$ and $Z$ bosons. $H$ stands for the
  physical Higgs boson.}
\label{diagrammi}
\end{figure}
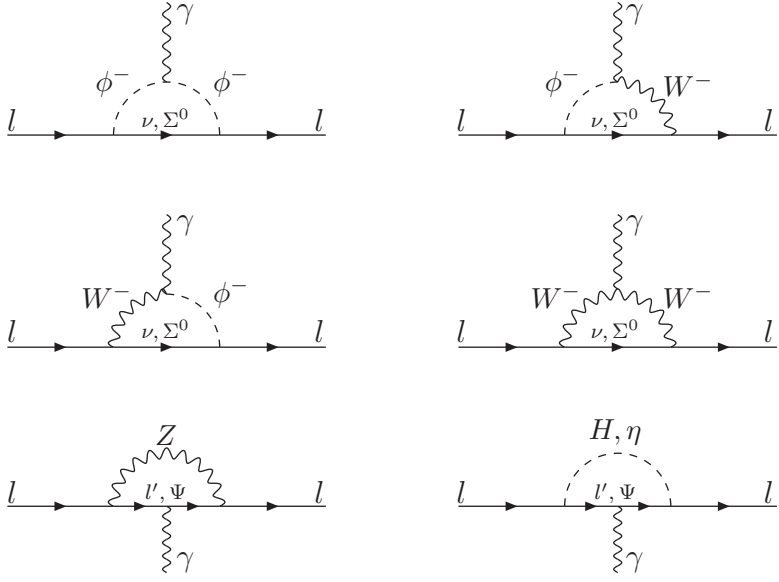

\subsection{Calculation of the anomalous magnetic moment}
\label{calc}

The anomalous magnetic moment of a lepton $l$ is related to the
photon-lepton vertex function $\Gamma^\mu$ as
follows~\cite{Heinemeyer:2004gx}:
\bea
\label{gamma-F}
\overline{u_l}(p-q)\Gamma^\mu u_l(p)&=&
\overline{u_l}(p-q)\left[\gamma^\mu F_V(q^2)+2p^\mu F_M(q^2)+\dots\right]u_l(p)\\
\label{F0-a}
a_l&=&-2m_lF_M(0)\, ,
\eea
where $p$ is the momentum of the incoming lepton and $q$ of the
outcoming photon.  We calculate the anomalous magnetic moment at
${\cal O}((Y_{\Sigma}v/M_\Sigma)^2)$, which is a good approximation as
long as $M_\Sigma$ is sufficiently big compared to
$Y_{\Sigma}v$~\footnote{$Y_{\Sigma}$ is ${\cal O}(1)$ or smaller. In
  particular, for the lowest allowed value of $M_\Sigma$, 100~GeV (see
  later), $Y_{\Sigma}\lesssim 10^{-2}$, to satisfy current EW bounds (see
  later).}.  The diagrams which contribute are depicted in
Fig.~1. Grouping the amplitudes according to the fermion circulating
in the loop, in the limit in which $m_l^2/M_W^2\to 0$, we have (see
Appendix~B for details):
\begin{eqnarray}
\label{Tnu}
T^{III}_{\nu}&=&\frac{i e G^{SM}_F}{8\sqrt{2}\pi^2}m_l\Big\{
\overline{u_l}(p-q)(2p\cdot\varepsilon)u_l(p)\Big\}
\left[\frac{5}{3}+\frac{7}{6}\epsilon_{ll}
-\frac{1}{2}\sum_i x_{\nu_{i}}V_{li}V_{il}^\dagger\right]\\
\label{Tmu}
T^{III}_l&=&\frac{i e G^{SM}_F}{8\sqrt{2}\pi^2}m_l\Big\{
\overline{u_l}(p-q)(2p\cdot\varepsilon)u_l(p)\Big\}\times\nn\\
&\times&\left[\frac{2}{3}-4\cos^2\theta_W+\frac{8}{3}\cos^4\theta_W
+\epsilon_{ll}\left(-\frac{16}{3}+\frac{8}{3}\cos^2\theta_W
\right)\right]\\
\label{TSigma}
T^{III}_\Sigma&=&\frac{i e G^{SM}_F}{8\sqrt{2}\pi^2}m_l\Big\{
\overline{u_l}(p-q)(2p\cdot\varepsilon)u_l(p)\Big\}
\sum_i \frac{v^2}{2}
\left(Y^{\dagger}_{\Sigma}M^{-1}_{\Sigma}\right)_{li}
\left(M^{-1}_{\Sigma}Y_{\Sigma}\right)_{il}
A(x_i)\\
\label{TPsi}
T^{III}_\Psi&=&\frac{i e G^{SM}_F}{8\sqrt{2}\pi^2}m_l\Big\{
\overline{u_l}(p-q)(2p\cdot\varepsilon)u_l(p)\Big\}
\sum_i \frac{v^2}{2}
\left(Y^{\dagger}_{\Sigma}M^{-1}_{\Sigma}\right)_{li}
\left(M^{-1}_{\Sigma}Y_{\Sigma}\right)_{il}\times\nn\\
&\times&\left[B(y_i)+C(z_i)\right]
\end{eqnarray}
where
\bea
A(x_i)&=&\frac{-38+185x_i-246x^2_i+107x^3_i
-8x^4_i+18(4-3x_i)x^2_i\log x_i}
{6(x_i-1)^4}\\
B(y_i)&=&\frac{40-46y_i-3y_i^2+2y_i^3+
7y_i^4+18(4-3y_i)y_i\log y_i}{6(y_i-1)^4}\\
C(z_i)&=& \frac{-16z_i+45z_i^2-36z_i^3
+7z_i^4+6(3z_i-2)z_i\log z_i}{6(z_i-1)^4}
\eea
and $x_{\nu_{i}}\equiv \frac{m^2_{\nu_i}}{M_W^2}$, $x_i\equiv
\frac{M^2_{\Sigma_i}}{M_W^2}$,
$y_i\equiv\frac{M^2_{\Sigma_i}}{M_Z^2}$,
$z_i\equiv\frac{M^2_{\Sigma_i}}{M_H^2}$. Moreover $V$ is the unitary
PMNS leptonic mixing matrix and $\epsilon_{ll}=\sum_i
\frac{v^2}{2}(Y_\Sigma^\dagger)_{li}M^{-2}_{\Sigma_i} (Y_\Sigma)_{il}$
is our expansion parameter~\footnote{Notice that $x_{\nu_{i}}\sim
  \mathcal{O}(\epsilon )$, as it can be easily seen looking at
  Eq.~(\ref{mnu}).}, not to be confused with $\varepsilon$ which
represents the polarization of the photon. $\epsilon$ coincides
with the coefficient of the unique dimension-six operator generated by
the fermionic triplets once they have been integrated out and it is
the unique combination of parameters which enters in all the low
energy processes, a part from neutrino masses.

Notice that the $\epsilon$ and $x_{\nu_{i}}$-independent terms in
Eq.~(\ref{Tnu})-(\ref{Tmu}) are precisely the 1-loop EW
contributions~\cite{EWg-2} and must be subtracted. The type-III seesaw
contribution to the anomalous magnetic moment is then:
\bea
\label{generaldeltaa}
\Delta a^{III}_l &=& \frac{G^{SM}_F m_l^2}{4\sqrt{2}\pi^2}\
\Bigg\{
-\frac{1}{2}\sum_i x_{\nu_{i}}V_{li}V_{il}^\dagger
+\sum_i \frac{v^2}{2}
\left(Y^{\dagger}_{\Sigma}M^{-1}_{\Sigma}\right)_{li}
\left(M^{-1}_{\Sigma}Y_{\Sigma}\right)_{il}\times\nn\\
&\times&\left[\frac{7}{6}-\frac{16}{3}+\frac{8}{3}\cos^2\theta_W
+A(x_i)+B(y_i)+C(z_i)\right]\Bigg\} \, .
\eea
%
%
%
The terms contained in the square parenthesis in the above expression
constitute the loop factor which turns out to be a negative function
which increases monotonically with $M_\Sigma$. Varying $M_\Sigma$ from
$100.8$~GeV to infinity~\footnote{We choose, as lower limit on
  $M_\Sigma$, the LEP bound on the mass of new heavy charged
  particles~\cite{PDG}, while the upper limit (infinity) is considered
  just in order to see which is the maximal value of the loop
  factor. In this case $\epsilon$ would go to zero, as well as
  $\Delta a^{III}_l$.}, it varies between -2.86 (-3.18) to -1.11,
taking $M_H=114.4$~GeV (250~GeV). Since $\sum_i
(Y^{\dagger}_{\Sigma}M^{-1}_{\Sigma})_{li}
(M^{-1}_{\Sigma}Y_{\Sigma})_{il}$ is a positive quantity, as well as
$\sum_i x_{\nu_{i}}V_{li}V_{il}^\dagger$, this implies that the
type-III seesaw contribution to the anomalous magnetic moment is
always negative. This already tells us that, even if we lower the
scale of the new physics, within this model we will not be able to
explain the measured discrepancy in the muon sector.

Omitting flavour indices, from the seesaw formula, Eq.~(\ref{mnu}),
one gets, in general, $\epsilon\sim m_\nu/M_\Sigma\sim 10^{-26} \,
(10^{15}\,\hbox{GeV}/M_\Sigma)$, where $m_\nu\sim \sqrt{\delta
  m^2_{atm}}$ has been used. On the other hand, since $x_\nu\sim\delta
m^2_{atm.}/M_W^2\sim 10^{-24}$, for large $M_\Sigma$ the first term in
Eq.~(\ref{generaldeltaa}) dominates and the contribution to the
anomalous magnetic moment of leptons is negligible. When this
relationship between $\epsilon$ and neutrino masses exists, even if
$M_\Sigma$ is lowered to the electroweak scale gaining thus several
orders of magnitudes, the type-III contribution will remain
negligible. However there are cases where $\epsilon$ can be decoupled
from neutrino masses or, in other words, where the smallness of these
is obtained through cancellations which do not affect
$\epsilon$~\cite{ABBGH,Kersten:2007vk}. In these cases the second term
in Eq.~(\ref{generaldeltaa}) is enhanced and the first one can be
neglected.  Since we want to find an upper bound on the size of the
total type-III contribution, we consider the case in which the term
proportional to $x_{\nu_i}$ can be neglected and $M_\Sigma$ is low.
In this case, and taking for the Higgs mass the lower limit
$M_H=114.4$~GeV, we have:
\bea
\label{asmallM}
\Delta a^{III}_l=\frac{G^{SM}_F m_l^2}{4\sqrt{2}\pi^2}(-2.86)\epsilon_{ll}\, .
\eea
What is now important in order to give an estimate of $\Delta a_l$ is
the value of $\epsilon_{ll}$. We will discuss this in the next
Section.

\subsection{The electron, muon and tau anomalous magnetic moments}
\label{aeamuatau}

The combination of the parameters of the type-III seesaw model which
enters into this calculation, \emph{i.\,e.} $\epsilon_{ll}$, is the
same that appears in many EW processes like leptonic and semileptonic
decays of the $W$ and $Z$ bosons. Considering these processes, upper
bounds on $\epsilon_{ll}$ have been derived in Ref.~\cite{ABBGH}:
$|\epsilon_{ee}|<3\cdot 10^{-3}$ and
$|\epsilon_{\mu\mu}|=|\epsilon_{\tau\tau}|<4\cdot 10^{-3}$. Taking
them into account, we obtain the following constraints on $\Delta
a^{III}_l$, in the case of ``small'' $M_\Sigma$ and with
$M_H=114.4$~GeV (250~GeV):
\bea
\label{ae}
|\Delta a^{III}_e|&<&4.66\ \ (5.18)\ \cdot 10^{-16}\\
\label{amu} 
|\Delta a^{III}_\mu|&<&2.67\ \ (2.97)\ \cdot 10^{-11}\\ 
\label{atau}
|\Delta a^{III}_\tau|&<&7.55\ \ (8.39)\ \cdot 10^{-9}\, .
\eea
We observe that the upper values obtained in
Eqs.~(\ref{ae})-(\ref{atau}) are smaller than the theoretical errors in
the SM computation, even if, for $\Delta a^{III}_e$ and $\Delta
a^{III}_\mu$, they are of the same order of magnitude as the error in
the EW part. Moreover, the value of $|\Delta a^{III}_\mu|$ obtained in
this model is two orders of magnitude smaller than the measured
discrepancy $\delta a_\mu$.

\section{The anomalous magnetic moment in the type-I seesaw model}
\label{typeI}

From the results obtained in the previous section it is now
straightforward to derive the contribution to the anomalous magnetic
moment in the type-I seesaw. Indeed in this case only singlet fermions
are present, which can be identified with the fields $\Sigma^0$ of the
type-III seesaw model. Consequently, among the diagrams depicted in
Fig.~1, only the ones containing $\nu$ and $\Sigma^0$ as internal
lines have to be considered.

The relevant couplings in this case are:
\be
\begin{array}{llll}
g^{CC}_{L_{l\nu}}=U_{0_{\nu\nu}}&
g^{CC}_{L_{l\Sigma}}=Y_{\Sigma}^{\dagger}M_{\Sigma}^{-1}\frac{v}{\sqrt{2}}&
g^{CC}_{R_{l\nu}}=0&
g^{CC}_{R_{l\Sigma}}=0\\
g^{\phi^{-}}_{L_{\nu}}=m_l U_{0_{\nu\nu}}&
g^{\phi^{-}}_{L_{\Sigma}}=m_lY_{\Sigma}^{\dagger}M_{\Sigma}^{-1}\frac{v}{\sqrt{2}}&
g^{\phi^{-}}_{R_{\nu}}=m_{\nu}^{*} U^{*}_{0_{\nu\nu}}&
g^{\phi^{-}}_{R_{\Sigma}}=Y_{\Sigma}^{\dagger}\frac{v}{\sqrt{2}}
\left(1-\frac{\epsilon'}{2}\right)
\end{array}\, .
\ee
Again, grouping the diagrams according to the fermion circulating in
the loop and making the same approximations we did in the previous
section, we obtain, at $\mathcal{O}(\epsilon)$:
\begin{eqnarray}
\label{TnuI}
T^I_{\nu}&=&\frac{i e G^{SM}_F}{8\sqrt{2}\pi^2}m_l\Big\{
\overline{u_l}(p-q)(2p\cdot\varepsilon)u_l(p)\Big\}
\left[\frac{5}{3}-\frac{5}{3}\epsilon_{ll}
-\frac{1}{6}\sum_i x_{\nu_{i}}V_{li}V_{il}^\dagger\right]\\
\label{TSigmaI}
T^I_\Sigma&=&\frac{i e G^{SM}_F}{8\sqrt{2}\pi^2}m_l\Big\{
\overline{u_l}(p-q)(2p\cdot\varepsilon)u_l(p)\Big\}
\sum_i \frac{v^2}{2}
\left(Y^{\dagger}_{\Sigma}M^{-1}_{\Sigma}\right)_{li}
\left(M^{-1}_{\Sigma}Y_{\Sigma}\right)_{il}
D(x_i)
\end{eqnarray}
where
\be
\label{D}
D(x_i)=\frac{10-55x_i+90x^2_i-37x^3_i
-8x^4_i+6(7x_i-4)x^2_i\log x_i}
{6(x_i-1)^4}\, .
\ee
In Eq.~(\ref{TnuI}) we recognize again the EW contribution of
neutrinos which we have to subtract. The contribution to the anomalous
magnetic moment of type-I seesaw is then:
\be
\label{generaldeltaaI}
\Delta a^{I}_l = \frac{G^{SM}_F m_l^2}{4\sqrt{2}\pi^2}
\left\{
\sum_i \frac{v^2}{2}
\left(Y^{\dagger}_{\Sigma}M^{-1}_{\Sigma}\right)_{li}
\left(M^{-1}_{\Sigma}Y_{\Sigma}\right)_{il}
\left[D(x_i)-\frac{5}{3}\right] 
-\frac{1}{6}\sum_i x_{\nu_{i}}V_{li}V_{il}^\dagger\right\}\, .
\ee
As before, we are interested in cases where the first term dominates
over the second, so that, from now on, we will neglect the latter.  The
terms contained in the square parenthesis in the above expression
constitute the loop factor which turns out to be a negative
monotonically-decreasing function. Then, also in this model, the
contribution to the anomalous magnetic moment of the leptons is always
negative and the measured deviation in the muon sector can not thus be
explained.

Also for the type-I seesaw, considering EW decays, it is possible to
put bounds on $\epsilon_{ll}$. These, which are slightly larger than
the ones obtained in the type-III case, $|\epsilon_{ll}|< 10^{-2}$,
have been derived in Ref.~\cite{uni}. Taking them into account and
considering the case in which $M_\Sigma=100$~GeV (1~TeV), we obtain,
for each lepton:
\bea
\label{aeI}
|\Delta a^{I}_e|&<& 6.18 \cdot 10^{-16}\quad (1.54 \cdot 10^{-15})\\
\label{amuI} 
|\Delta a^{I}_\mu|&<& 2.65 \cdot 10^{-11}\quad (6.63 \cdot 10^{-11})\\ 
\label{atauI}
|\Delta a^{I}_\tau|&<& 7.50 \cdot 10^{-9}\quad\,\, (1.87 \cdot 10^{-8})\, .
\eea
We observe that also in this model the new contributions are
generically smaller than the SM error. However, if we consider
$|\Delta a^{I}_\tau|$, for masses of order of 1~TeV or higher, we see
that it can be comparable with the present SM error.

\section{Conclusions}
\label{conclu}

Motivated by the deviation from the SM value of the measured anomalous
magnetic moment of the muon and considering that new physics beyond
the SM is required to explain neutrino masses, we have calculated the
contribution to the anomalous magnetic moment of the leptons in two
classes of models for generating neutrino masses, \emph{i.\,e.} the
type-I and type-III seesaw models.

We have found that, even if the scale of such new physics is just
behind the corner, in the reach of future accelerators like LHC, its
contribution to the anomalous magnetic moment of the leptons is
generically smaller than the theoretical errors in the SM calculation,
rendering thus impossible to distinguish from the SM.  Notice that
what forces this contribution to be so small are the bounds on
$\epsilon_{ll}$ coming from other EW processes. Since in seesaw models
like these the combination of the parameters of the new physics which
enters into low energy processes is unique, the strong bounds coming
from various EW processes apply, making impossible to use this new
physics to explain the discrepancy of the muon anomalous magnetic
moment. This is not the case for example in supersymmetric
theories~\cite{Stockinger:2006zn}, where the much larger parameters
space makes it possible that different combinations of them affect
different processes, permitting thus to evade EW bounds.

We can thus conclude that, even if new physics responsible for
neutrino mass generation like fermionic singlets or triplets will be
discovered at low scale, its contribution to the anomalous magnetic
moment of the leptons will anyway be extremely small.

\section*{Note added}

During the completion of this work, Ref.~\cite{Chao:2008iw} appeared,
calculating the anomalous magnetic moment of the muon in three
low-scale seesaw models. Our results differ from the ones presented in
that paper in the case of type-III seesaw by one order of
magnitude~\footnote{A small difference is also present in the type-I
  case, due to an extra factor 2 in the diagram containing two
  Goldstone bosons circulating in the loop.}. Firstly, in
Ref.~\cite{Chao:2008iw} only the Higgs diagram is calculated, arguing
that this is the relevant one. However, from our
Eqs.~(\ref{Tnu})-(\ref{generaldeltaa}) it can be seen that all the
diagrams give a contribution of the same order of magnitude. This is a
difference, but it does not lead to one order of magnitude of
discrepancy. This discrepancy comes from the fact that in
Ref.~\cite{Chao:2008iw} existing bounds on $\epsilon_{ll}$ have not
been taken into account. If we take the result of
Ref.~\cite{Chao:2008iw}, Eqs.~(20)-(21), re-express it in terms of
$\epsilon$ and take the limit for small $M_\Sigma$, we obtain a formula
similar to our Eq.~(\ref{asmallM}), but with the numerical factor
being even one order of magnitude smaller. So it seems to be
impossible to get a contribution of the order of $10^{-10}$, as
claimed in that paper, if the existing bounds are taken into account.

\section*{Acknowledgments}

We acknowledge discussions with S.~B\'ejar, E.~Fernand\'ez-Mart\'\i nez,
T.~Hambye, A.~Ibarra, M.~Passera, S.~Rigolin and S.~Vempati. We
especially thank M.~Papucci for useful discussions and for precious
comments on the first version of this manuscript.

\section*{Appendix A}

The Lagrangian terms in Eq.~(\ref{fulllagrangian}), which are the ones
relevant for our calculation, are:
\bea
\label{CC}
\mathcal{L}_{CC}&=&\frac{g}{\sqrt{2}}\left(\begin{array}{cc}\overline{l} 
& \overline{\Psi}\end{array}\right)\gamma^{\mu}W^{-}_{\mu}
\left(P_L g^{CC}_L+P_R g^{CC}_R\sqrt{2}\right)
\left(\begin{array}{c}\nu \\ \Sigma^0\end{array}\right)+\textrm{h.c}.\nn\\
%
\label{NC}
\mathcal{L}_{NC}&=&\frac{g}{cos\theta_W}\left(\begin{array}{cc}\overline{l} & \overline{\Psi}\end{array}\right)\gamma^{\mu}Z_{\mu}\left(P_L g^{NC}_L+P_R g^{NC}_R\right)\left(\begin{array}{c}l \\ \Psi\end{array}\right)\nn\\
%
\label{Heta}
\mathcal{L}_{H,\eta}&=&\frac{g}{2M_W}\left(\begin{array}{cc}\overline{l} & \overline{\Psi}\end{array}\right)H\left(P_L g^{H}_L+P_R g^{H}_R\right)\left(\begin{array}{c}l \\ \Psi\end{array}\right)\nn\\
&+&i\frac{g}{2M_W}\left(\begin{array}{cc}\overline{l} & \overline{\Psi}\end{array}\right)\eta\left(P_L g^{\eta}_L+P_R g^{\eta}_R\right)\left(\begin{array}{c}l \\ \Psi\end{array}\right)\nn\\
\label{phi}
\mathcal{L}_{\phi^{-}}&=&-\phi^{-}\overline{l}\frac{g}{\sqrt{2}M_W}
\left\{\left(P_L g^{\phi^{-}}_{L_{\nu}}+P_R g^{\phi^{-}}_{R_{\nu}}\right)\nu
+\left(P_L g^{\phi^{-}}_{L_{\Sigma}}+P_R g^{\phi^{-}}_{R_{\Sigma}}\right)
\Sigma^0\right\}+\textrm{h.c.}\nn
\eea
with
\bea
\label{cc}
g^{CC}_{L}&=&\left(\begin{array}{cc}
g^{CC}_{L_{l\nu}}&g^{CC}_{L_{l\Sigma}}\\
g^{CC}_{L_{\Psi\nu}}&g^{CC}_{L_{\Psi\Sigma}}\end{array}\right)
= \left(\begin{array}{cc}
\left(1+\epsilon\right)U_{0_{\nu\nu}} & 
            -Y_{\Sigma}^{\dagger}M_{\Sigma}^{-1}\frac{v}{\sqrt{2}}\\
0 & \dots \end{array}\right)\nn\\
g^{CC}_{R}&=&\left(\begin{array}{cc}
g^{CC}_{R_{l\nu}}&g^{CC}_{R_{l\Sigma}}\\
g^{CC}_{R_{\Psi\nu}}&g^{CC}_{R_{\Psi\Sigma}}\end{array}\right)= 
\left(\begin{array}{cc}
0 & -m_lY_{\Sigma}^{\dagger}M_{\Sigma}^{-2}v\\
-M_{\Sigma}^{-1}Y^{*}_{\Sigma}U^{*}_{0_{\nu\nu}}\frac{v}{\sqrt{2}} & \dots
\end{array}\right) \nn\\
g^{NC}_{L}&=&\left(\begin{array}{cc}
g^{NC}_{L_{ll}}&g^{NC}_{L_{l\Psi}}\\
g^{NC}_{L_{\Psi l}}&g^{NC}_{L_{\Psi\Psi}}\end{array}\right)= 
\left(\begin{array}{cc}
\frac{1}{2}-cos^2\theta_W-\epsilon & 
      \frac{1}{2}Y_{\Sigma}^{\dagger}M_{\Sigma}^{-1}v\\
\frac{1}{2}M_{\Sigma}^{-1}Y_{\Sigma}v & \dots \end{array}\right)\nn\\
g^{NC}_{R}&=&\left(\begin{array}{cc}
g^{NC}_{R_{ll}}&g^{NC}_{R_{l\Psi}}\\
g^{NC}_{R_{\Psi l}}&g^{NC}_{R_{\Psi\Psi}}\end{array}\right)= 
\left(\begin{array}{cc}
1-cos^2\theta_W & m_lY_{\Sigma}^{\dagger}M_{\Sigma}^{-2}v\\
 M_{\Sigma}^{-2}Y_{\Sigma}m_lv & \dots \end{array}\right)\nn\\
g^{H}_{L}&=&\left(\begin{array}{cc}
g^{H}_{L_{ll}}&g^{H}_{L_{l\Psi}}\\
g^{H}_{L_{\Psi l}}&g^{H}_{L_{\Psi\Psi}}\end{array}\right)= 
\left(\begin{array}{cc}
m_l\left(3\epsilon-1\right) & -m_lY_{\Sigma}^{\dagger}M_{\Sigma}^{-1}v\\
-Y_{\Sigma}v\left(1-\epsilon\right)-M_{\Sigma}^{-2}Y_{\Sigma}m_l^2v & 
                          \dots \end{array}\right)\nn\\
g^{H}_{R}&=&\left(\begin{array}{cc}
g^{H}_{R_{ll}}&g^{H}_{R_{l\Psi}}\\
g^{H}_{R_{\Psi l}}&g^{H}_{R_{\Psi\Psi}}\end{array}\right)= 
\left(\begin{array}{cc}
\left(3\epsilon-1\right)m_l & 
-\left(1-\epsilon\right)Y_{\Sigma}^{\dagger}v-m_l^2Y_{\Sigma}^{\dagger}M_{\Sigma}^{-2}v\\
-M_{\Sigma}^{-1}Y_{\Sigma}m_lv & \dots \end{array}\right)\nn\\
g^{\eta}_{R}&=&\left(\begin{array}{cc}
g^{\eta}_{R_{ll}}&g^{\eta}_{R_{l\Psi}}\\
g^{\eta}_{R_{\Psi l}}&g^{\eta}_{R_{\Psi\Psi}}\end{array}\right)= 
\left(\begin{array}{cc}
-\left(\epsilon+1\right)m_l & 
\left(1-\epsilon\right)Y_{\Sigma}^{\dagger}v-m_l^2Y_{\Sigma}^{\dagger}M_{\Sigma}^{-2}v\\
-M_{\Sigma}^{-1}Y_{\Sigma}m_lv & \dots \end{array}\right)\nn\\
\label{getal}
g^{\eta}_{L}&=&\left(\begin{array}{cc}
g^{\eta}_{L_{ll}}&g^{\eta}_{L_{l\Psi}}\\
g^{\eta}_{L_{\Psi l}}&g^{\eta}_{L_{\Psi\Psi}}\end{array}\right)= 
\left(\begin{array}{cc}
m_l\left(\epsilon+1\right) & m_lY_{\Sigma}^{\dagger}M_{\Sigma}^{-1}v\\
-Y_{\Sigma}v\left(1-\epsilon\right)+M_{\Sigma}^{-2}Y_{\Sigma}m_l^2v & 
            \dots \end{array}\right)\nn
\eea
and
\bea
\left\{\begin{array}{l}
g^{\phi^{-}}_{L_{\nu}}=m_l U_{0_{\nu\nu}}\\
g^{\phi^{-}}_{R_{\nu}}=-\left(1-\epsilon\right)m_{\nu}^{*} U^{*}_{0_{\nu\nu}}\end{array}\right. \quad
\left\{\begin{array}{l}
g^{\phi^{-}}_{L_{\Sigma}}=m_lY_{\Sigma}^{\dagger}M_{\Sigma}^{-1}\frac{v}{\sqrt{2}}\\
g^{\phi^{-}}_{R_{\Sigma}}=\left(1-\epsilon\right)Y_{\Sigma}^{\dagger}\frac{v}{\sqrt{2}}\left(1-\frac{\epsilon'^{*}}{2}\right)-\sqrt{2}m_{\nu}^{*}Y_{\Sigma}^{T}M_{\Sigma}^{-1}v\end{array}\right. \, .\nn
\eea
The dots in the previous equations refer to $\Sigma^0$-$\Sigma^0$ and
$\Psi$-$\Psi$ interactions which do not contribute to the process we
are considering. Moreover $\epsilon=\frac{v^2}{2}Y_\Sigma^\dagger
M^{-2}_\Sigma Y_\Sigma$, $\epsilon'=\frac{v^2}{2}M^{-1}_\Sigma
Y_\Sigma Y_\Sigma^\dagger M^{-1}_\Sigma$ and
$U_{0\nu\nu}=(1-\frac{\epsilon}{2}) V$, where $V$ is the lowest order
neutrino mixing matrix (PMNS matrix) which is unitary; finally we take
$\sin^2\theta_W=0.23$.

\section*{Appendix B}

Performing the calculation in the 't Hooft-Feynman gauge, after loop
integration, the various amplitudes, at ${\cal O}(\epsilon)$ and
neglecting terms proportional to $x_l=m_l^2/M_W^2$, $y_l=m_l^2/M_Z^2$,
$z_l=m_l^2/M_H^2$ and $m_l^2/M_\Sigma^2$ that will give irrelevant
contributions, are:
\begin{eqnarray}
T^{\phi^-/W^-}_{\nu_{i}}&=&\frac{i e G^{SM}_F}{8\sqrt{2}\pi^2}m_l\Big\{
\overline{u_l}(p-q)P_R(2p\cdot\varepsilon)u_l(p)\Big\}
\left[\left(1+\frac{\epsilon}{2}\right)_{l\alpha}V_{\alpha i}
V^\dagger_{i\beta}\left(1-\frac{\epsilon}{2}\right)_{\beta l}\right]
F_1(x_{\nu_i})\nn\\
T^{W^-/\phi^-}_{\nu_{i}}&=&\frac{i e G^{SM}_F}{8\sqrt{2}\pi^2}m_l\Big\{
\overline{u_l}(p-q)P_L(2p\cdot\varepsilon)u_l(p)\Big\}
\left[\left(1-\frac{\epsilon}{2}\right)_{l\alpha}V_{\alpha i}
V^\dagger_{i\beta}\left(1+\frac{\epsilon}{2}\right)_{\beta l}\right]
F_1(x_{\nu_i})\nn\\
T^{W^-/W^-}_{\nu_{i}}&=&\frac{i e G^{SM}_F}{8\sqrt{2}\pi^2}m_l\Big\{
\overline{u_l}(p-q)(2p\cdot\varepsilon)u_l(p)\Big\}
\left[\left(1+\frac{\epsilon}{2}\right)_{l\alpha}V_{\alpha i}
V^\dagger_{i\beta}\left(1+\frac{\epsilon}{2}\right)_{\beta l}\right]
F_2(x_{\nu_i})\nn\\
T^{\phi^-/\phi^-}_{\nu_{i}}&=&\frac{i e G^{SM}_F}{8\sqrt{2}\pi^2}m_l\Big\{
\overline{u_l}(p-q)(2p\cdot\varepsilon)u_l(p)\Big\}
\left[\left(1-\frac{\epsilon}{2}\right)_{l\alpha}V_{\alpha i}
V^\dagger_{i\beta}\left(1-\frac{\epsilon}{2}\right)_{\beta l}\right]
x_{\nu_{i}}F_3(x_{\nu_i})\nn
\eea
\bea
T^{\phi^-/W^-}_{\Sigma_{i}}&=&-\frac{i e G^{SM}_F}{8\sqrt{2}\pi^2}m_l\Big\{
\overline{u_l}(p-q)(2P_L+P_R)(2p\cdot\varepsilon)u_l(p)\Big\}
\frac{v^2}{2}
\left(Y^{\dagger}_{\Sigma}M^{-1}_{\Sigma}\right)_{li}
\left(M^{-1}_{\Sigma}Y_{\Sigma}\right)_{il} F_1(x_i)\nn\\
T^{W^-/\phi^-}_{\Sigma_{i}}&=&-\frac{i e G^{SM}_F}{8\sqrt{2}\pi^2}m_l\Big\{
\overline{u_l}(p-q)(P_L+2P_R)(2p\cdot\varepsilon)u_l(p)\Big\}
\frac{v^2}{2}
\left(Y^{\dagger}_{\Sigma}M^{-1}_{\Sigma}\right)_{li}
\left(M^{-1}_{\Sigma}Y_{\Sigma}\right)_{il} F_1(x_i)\nn\\
T^{W^-/W^-}_{\Sigma_{i}}&=&\frac{i e G^{SM}_F}{8\sqrt{2}\pi^2}m_l\Big\{
\overline{u_l}(p-q)(2p\cdot\varepsilon)u_l(p)\Big\}
\frac{v^2}{2}
\left(Y^{\dagger}_{\Sigma}M^{-1}_{\Sigma}\right)_{li}
\left(M^{-1}_{\Sigma}Y_{\Sigma}\right)_{il}
F_4(x_i)\nn\\
T^{\phi^-/\phi^-}_{\Sigma_{i}}&=&\frac{i e G^{SM}_F}{8\sqrt{2}\pi^2}m_l\Big\{
\overline{u_l}(p-q)(2p\cdot\varepsilon)u_l(p)\Big\}
\frac{v^2}{2}
\left(Y^{\dagger}_{\Sigma}M^{-1}_{\Sigma}\right)_{li}
\left(M^{-1}_{\Sigma}Y_{\Sigma}\right)_{il}
x_i F_5(x_i)\nn
\eea
\bea
T^{Z}_{\Psi_i}&=&\frac{i e G^{SM}_F}{8\sqrt{2}\pi^2}m_l\Big\{
\overline{u_l}(p-q)(2p\cdot\varepsilon)u_l(p)\Big\}
\frac{v^2}{2}
\left(Y^{\dagger}_{\Sigma}M^{-1}_{\Sigma}\right)_{li}
\left(M^{-1}_{\Sigma}Y_{\Sigma}\right)_{il}
2G_1(y_i)\nn\\
T^{H}_{\Psi_i}&=&\frac{i e G^{SM}_F}{8\sqrt{2}\pi^2}m_l\Big\{
\overline{u_l}(p-q)(2p\cdot\varepsilon)u_l(p)\Big\}
\frac{v^2}{2}
\left(Y^{\dagger}_{\Sigma}M^{-1}_{\Sigma}\right)_{li}
\left(M^{-1}_{\Sigma}Y_{\Sigma}\right)_{il}
y_iG_2(y_i)\nn\\
T^{\eta}_{\Psi_i}&=&\frac{i e G^{SM}_F}{8\sqrt{2}\pi^2}m_l\Big\{
\overline{u_l}(p-q)(2p\cdot\varepsilon)u_l(p)\Big\}
\frac{v^2}{2}
\left(Y^{\dagger}_{\Sigma}M^{-1}_{\Sigma}\right)_{li}
\left(M^{-1}_{\Sigma}Y_{\Sigma}\right)_{il}
z_iG_2(z_i)\nn
\eea
\bea
T^{Z}_{l'_i}&=&\frac{i e G^{SM}_F}{8\sqrt{2}\pi^2}m_l\Big\{
\overline{u_l}(p-q)(2p\cdot\varepsilon)u_l(p)\Big\}
\delta_{l'_i l}\Big[\left(G_4(y_{l'_i})-12\cos^2\theta_W G_3(y_{l'_i})\right.\nn\\
&&\left.+8\cos^4\theta_W G_3(y_{l'_i})\right)+
\epsilon_{ll}\left(-4G_5(y_{l'_i})+8\cos^2\theta_W G_3(y_{l'_i})\right)
\Big]\nn\\
T^{H}_{l'_i}&=&0\nn\\
T^{\eta}_{l'_i}&=&0\nn
\eea
where $x_{\nu_{i}}\equiv \frac{m^2_{\nu_i}}{M_W^2}$,
$x_i\equiv \frac{M^2_{\Sigma_i}}{M_W^2}$,
$y_{l'_i}\equiv\frac{m^2_{l'_i}}{M_Z^2}$,
$y_i\equiv\frac{M^2_{\Sigma_i}}{M_Z^2}$,
$z_i\equiv\frac{M^2_{\Sigma_i}}{M_H^2}$ and $F_i(x)$ and $G_i(x)$
are the following functions:
\begin{eqnarray}
\label{F1}
F_1(x)&=&J_0(x)-2J_1(x)+J_2(x)\nn\\
F_2(x)&=&3J_0(x)-8J_1(x)+7J_2(x)-2J_3(x)\nn\\
F_3(x)&=&J_1(x)-J_3(x)\nn\\
F_4(x)&=&-9J_0(x)+16J_1(x)-5J_2(x)-2J_3(x)\nn\\
F_5(x)&=&-3J_1(x)+4J_2(x)-J_3(x)\nn\\
G_1(x)&=&7I_1(x)-8I_2(x)+I_3(x)\nn\\
G_2(x)&=&2I_0(x)-3I_1(x)+I_3(x)\nn\\
G_3(x)&=&I_1(x)-2I_2(x)+I_3(x)\nn\\
G_4(x)&=&3I_1(x)-8I_2(x)+5I_3(x)\nn\\
G_5(x)&=&3I_1(x)-4I_2(x)+I_3(x)\nn
\eea
where~\footnote{The terms involving $x_l=m_l^2/M_W^2$ and
  $y_l=m_l^2/M_Z^2$ in the denominators of $J_n$ and $I_n$
  have been neglected.}
\be
J_n(x)=\int_0^1 \textrm{d}\alpha \frac{\alpha^n}{1-\alpha(1-x)}\hspace{1cm}
I_n(x)=\int_0^1 \textrm{d}\alpha \frac{\alpha^n}{x+\alpha(1-x)}\, .\nn
\ee
In the limit in which $x_{\nu_{i}}\to 0$ and $y_l\to 0$, taking the
linear terms in $x_{\nu_{i}}$, we have:
\be
\begin{array}{lll}
F_1(x_{\nu_{i}})\to \frac{1}{2}-\frac{1}{2}x_{\nu_{i}} &
F_2(x_{\nu_{i}})\to \frac{7}{6}-\frac{5}{6}x_{\nu_{i}} &
F_3(x_{\nu_{i}})\to \frac{5}{6}+\frac{10}{3}x_{\nu_{i}}\\[.2cm]
G_3(y_l)\to \frac{1}{3} &
G_4(y_l)\to \frac{2}{3} &
G_5(y_l)\to \frac{4}{3}\, .
\end{array}\nn
\ee
In this limit, grouping the various amplitudes according to the
fermion circulating in the loop and summing over $i$, we get the
results displayed in the text, Eqs.~(\ref{Tnu})-(\ref{TPsi}).


\end{document}